\documentclass[a4paper,11pt]{article}
\usepackage{pos}

\title{Bottomonium suppression in pNRQCD and open quantum system approach}
\ShortTitle{Bottomonium suppression in pNRQCD and open quantum system approach}

\author*[a,b]{Ajaharul Islam}
\author[c]{Nora Brambilla}
\author[d]{Miguel \'{A}ngel Escobedo}
\author{Michael Strickland}
\author[c]{Antonio Vairo}
\author[e,f]{Peter Vander Griend}

\affiliation[a]{Department of Physics, Kent State University, Kent, OH 44242, United States}

\affiliation[b]{Institute of Particle Physics and Key Laboratory of Quark and Lepton Physics (MOE), Central China Normal University, Wuhan, 430079, China}

\affiliation[c]{Physik-Department, Technische Universit\"{a}t M\"{u}nchen, James-Franck-Str.~1, 85748 Garching,
Germany}

\affiliation[d]{Departament de Física Quàntica i Astrofísica and Institut de Ciències del Cosmos, Universitat de Barcelona, Martí i Franquès 1, 08028 Barcelona, Spain}

\affiliation[e]{Department of Physics and Astronomy, University of Kentucky, Lexington, KY 40506, United States}

\affiliation[f]{Theoretical Physics Department, Fermilab, P.O. Box 500, Batavia, IL 60510, United States}

\emailAdd{nora.brambilla@tum.de}
\emailAdd{miguel.a.escobedo@fqa.ub.edu}
\emailAdd{aislam2@kent.edu}
\emailAdd{michael.strickland@gmail.com}
\emailAdd{antonio.vairo@tum.de}
\emailAdd{peter.vandergriend@uky.edu}

\abstract{
By employing the potential non-relativistic quantum chromodynamics (pNRQCD) effective field theory within an open quantum system framework, we derive a Lindblad equation governing the evolution of the heavy-quarkonium reduced density matrix, accurate to next-to-leading order (NLO) in the ratio of the state's binding energy to the medium's temperature \cite{Brambilla:2022ynh}. The derived NLO Lindblad equation provides a more reliable description of heavy-quarkonium evolution in the quark-gluon plasma at low temperatures compared to the leading-order truncation. For phenomenological applications, we numerically solve this equation using the quantum trajectories algorithm. By averaging over Monte Carlo-sampled quantum jumps, we obtain solutions without truncation in the angular momentum quantum number of the considered states. Our analysis highlights the importance of quantum jumps in the nonequilibrium evolution of bottomonium states within the quark-gluon plasma \cite{Brambilla:2023hkw}. Additionally, we demonstrate that the quantum regeneration of singlet states from octet configurations is essential to explain experimental observations of bottomonium suppression. The heavy-quarkonium transport coefficients used in our study align with recent lattice QCD determinations.}

\FullConference{The XVIth Quark Confinement and the Hadron Spectrum Conference (QCHSC24)\\
 19-24 August, 2024\\
 Cairns Convention Centre, Cairns, Queensland, Australia\\}

\begin{document}
\maketitle

\section{Introduction}
Matsui and Satz \cite{Matsui:1986dk} first proposed heavy-quarkonium suppression as a signal of quark-gluon plasma (QGP) formation, making it a key focus of heavy-ion experiments ever since. Laine et al. \cite{Laine:2006ns} showed in their perturbative calculation that, in addition to the Debye screening of the real part, the heavy-quark potential also contains a significant imaginary part. The imaginary part of the potential, driven by Landau damping \cite{Brambilla:2013dpa} and singlet-to-octet transitions \cite{Brambilla:2008cx}, gives quarkonium a medium-induced decay width, leading to finite lifetime, in-medium dissolution, and suppression. For an accurate description of in-medium quarkonium evolution which must account for the thermal width, screening effects, and the recombination process; we adopt the Open Quantum System (OQS) approach. In this framework, heavy quarkonium is treated as an open quantum system interacting with a medium of light quarks and gluons. The key quantity is the reduced density matrix, derived by tracing out the environment's degrees of freedom from the full density matrix. The resulting equation is called the master equation which describes the evolution of the reduced density matrix \cite{Akamatsu:2011se, Brambilla:2016wgg, Sharma:2019xum, Blaizot:2015hya, Yao:2018nmy, Katz:2015qja}. In this article, we will describe our work \cite{Brambilla:2022ynh, Brambilla:2023hkw} on an approach combining the OQS framework with the potential NRQCD (pNRQCD) to exploit the evolution of heavy-quarkonium inside the QGP.

This article is organized as follows. In Section \ref{sec2}, we give an idea about how heavy-quarkonium is considered as an open quantum system in pNRQCD. In Section \ref{sec3}, we represent the pNRQCD master equation. In Section \ref{sec4}, we obtain the LO Lindblad equation in E/T expansion. In Section \ref{sec5}, we obtain the NLO Lindblad equation in E/T expansion. In Section \ref{sec6}, we obtain jump operators, width operators, and the effective Hamiltonian in the reduced spherical representation. We describe the QTraj algorithm, phenomenological results, summary and conclusion in Sections \ref{sec7}, \ref{sec8}, and \ref{sec9}, respectively.

\section{Heavy quarkonium as an Open Quantum System in pNRQCD}
\label{sec2}
\begin{figure}[ht]
\begin{center}
\includegraphics[width=0.35\linewidth]{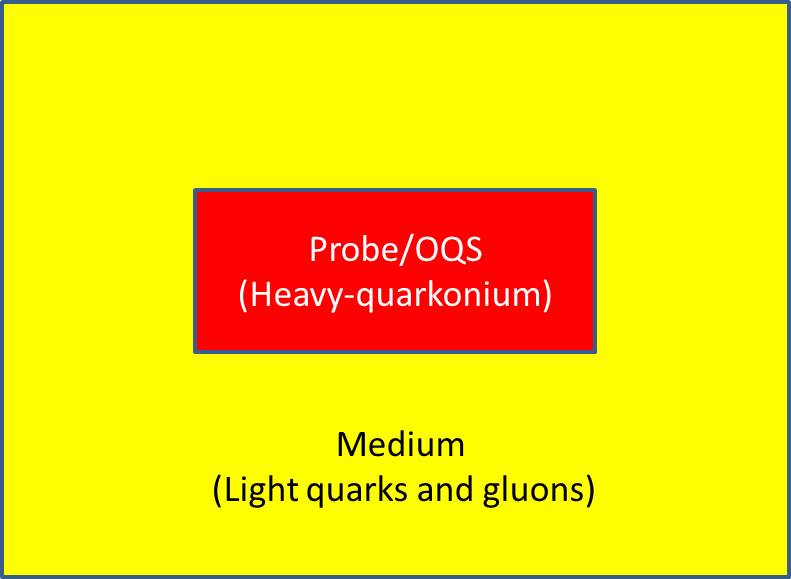} \hspace{3mm}
\includegraphics[width=0.52\linewidth]{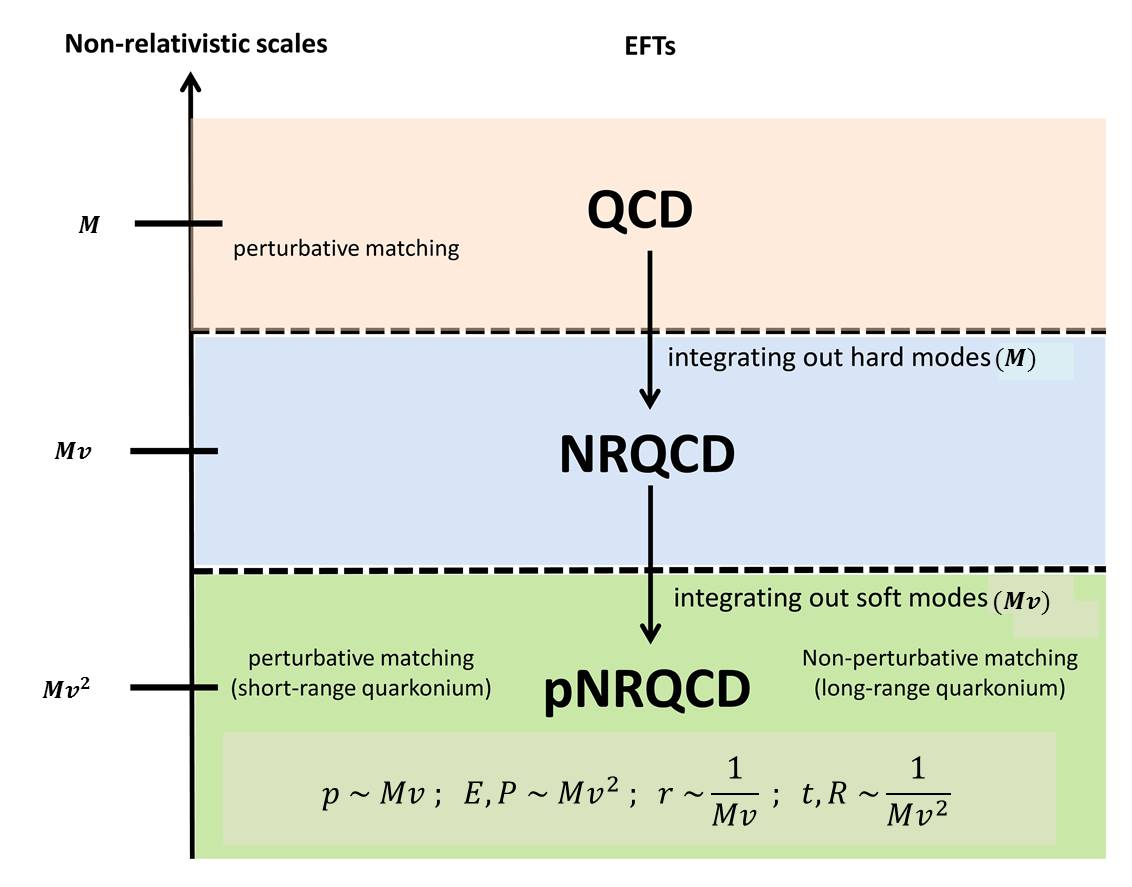}
\end{center}
\caption{Left panel: Heavy-quarkonium as an open quantum system inside QGP. Right panel: Hierarchy of energy scales and effective field theories (EFTs) for systems made of a heavy quark and antiquark pair \cite{Gross:2022hyw}.
}
\label{fig01}
\end{figure}

In effective field theories (EFTs), heavy-quarkonium states are characterized by the heavy quark mass $M$, and the non-relativistic relative velocity $v\ll 1$. These states follow the following hierarchy of scales: $M~(\mathrm{hard~modes})\gg M v~(\mathrm{soft~modes})\gg M v^2~(\mathrm{ultrasoft~modes})$ . We know integrating out the hard scale from  QCD gives rise to non-relativistic QCD (NRQCD) whereas integrating out the soft scale gives rise to potential NRQCD (pNRQCD) (see Fig.~\ref{fig01}).
The degrees of freedom in the pNRQCD Lagrangian are heavy-heavy bound states in color singlet and color octet configurations and gluons and light quarks at the ultrasoft scale. Inside QGP, the heavy-quarkonium state is considered as the open quantum system while light quarks and gluons as the medium. The Hamiltonian for the total system (open quantum system and medium) is given by
\begin{equation}
H_{\rm total}=H_{\rm sys}\otimes I_{\rm med} + I_{\rm sys}\otimes H_{\rm med} + H_{\rm int}\,\, , 
\label{eq:01}
\end{equation}
and the time evolution of the total density matrix is given by
\begin{equation}
\frac{d}{dt}\rho_{\rm total}(t)=\frac{1}{i\hbar}\big[H_{\rm total},\rho_{\rm total}(t)\big]\,\, ,
\label{eq:02}
\end{equation}

where $H_{int}$ represents the interaction between the heavy quarkonium and the medium. 
Starting with the initial condition, $\rho_{\rm total} (0)=\rho_{\rm sys}\otimes \rho_{\rm med}$, we integrate out the degrees of freedom that correspond to the medium and obtain the reduced density matrix of the system, $\rho_{\rm sys}(t)\equiv Tr_{\rm med}\big\{\rho_{\rm total}(t)\big\}$.

\begin{figure}[ht]
\begin{center}
\includegraphics[width=0.45\linewidth]{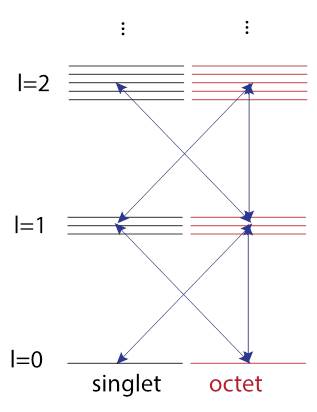}
\end{center}
\caption{Transitions between singlet and octet states.
}
\label{fig1.2}
\end{figure}
In the end, we obtain the time evolution for the reduced density matrix $\rho_{\rm sys}$ in Lindblad form, which is given by
\begin{equation}
    \frac{d \rho_{\rm sys}}{dt} = -i \big[H_{\rm sys}, \rho_{\rm sys}\big] + \sum_n \Big(C_n ~\rho_{\rm sys}~ C^\dagger_n - \frac{1}{2}\big\{C^\dagger_n C_n, \rho_{\rm sys}\big\}\Big)\,\, ,
\end{equation}
where $C_n$ are known as jump or collapse operators. We define partial, total widths as, $ \Gamma_n = C^\dagger_n C_n,\quad\Gamma = \sum_n \Gamma_n$, and a non-Hermitian effective Hamiltonian as, $H_{\rm eff} = H_{\rm sys} - \frac{i}{2} \Gamma$, to rewrite the Lindblad equation as
\begin{equation}
    \frac{d \rho_{\rm sys}}{dt} = -i H_{\rm eff}~ \rho_{\rm sys} + i\rho_{\rm sys} ~H_{\rm eff}^\dagger + \sum_n C_n ~\rho_{\rm sys}~ C^\dagger_n\,\, .
\end{equation}
In the above equation, the term $C_n ~\rho_{\rm sys}~ C^\dagger_n$ is known as the jump term, which is responsible for the transitions between different singlet and octet states.\\

Fig~\ref{fig1.2} describes transitions among different quantum states. A singlet state can absorb a gluon and become an octet state: ${\rm singlet} (s) + {\rm gluon} (g)\rightarrow {\rm octet} (o)$, whereas an octet state can emit a gluon and become a singlet state: ${\rm octet} (o) \rightarrow {\rm singlet}(s) + {\rm gluon} (g) $.

\section{The pNRQCD master equation}
\label{sec3}
If we do not make any assumption on the relation between $T$ and $E$, we have the evolution of the reduced density matrix of heavy quarkonium ~\cite{Brambilla:2016wgg,Brambilla:2017zei}.
\begin{align}
	\frac{d\rho_{s}(t)}{dt} &= -i[h_{s}, \rho_{s}(t)] - \Sigma_{s} \rho_{s}(t) - \rho_{s}(t) \Sigma^{\dagger}_{s} + \Xi_{so}(\rho_{o}(t)),\\
	\frac{d\rho_{o}(t)}{dt} &= -i[h_{o}, \rho_{o}(t)] - \Sigma_{o} \rho_{o}(t) - \rho_{o}(t) \Sigma_{o}^{\dagger}
	+ \Xi_{os}(\rho_{s}(t)) + \Xi_{oo}(\rho_{o}(t)),
\end{align}
\begin{align}
	\Sigma_{s} &= r_{i} A_{i}^{so \dagger}, \\
	\Sigma_{o} &= \frac{1}{N_{c}^{2}-1} r_{i} A_{i}^{os \dagger}
	+ \frac{N_{c}^{2}-4}{2(N_{c}^{2}-1)}r_{i}A_{i}^{oo \dagger}, \\
	\Xi_{so}(\rho_{o}(t)) &= \frac{1}{N_{c}^{2}-1} \left( A_{i}^{os \dagger} \rho_{o}(t) r_{i} + r_{i} \rho_{o}(t) A_{i}^{os}\right), \\
	\Xi_{os}(\rho_{s}(t)) &= A_{i}^{so \dagger} \rho_{s}(t) r_{i} 
	+ r_{i} \rho_{s}(t) A_{i}^{so}, \\
	\Xi_{oo}(\rho_{o}(t)) &= \frac{N_{c}^{2}-4}{2(N_{c}^{2}-1)}\left(
	A_{i}^{oo \dagger} \rho_{o}(t) r_{i} + r_{i} \rho_{o}(t) A_{i}^{oo}\right),
\end{align}
with
\begin{align}
	A_{i}^{uv} &= \frac{g^{2}}{6N_{c}} \int^{\infty}_{0} \text{d}s\, e^{-i h_{u}s} r_{i} e^{i h_{v} s} 
	\langle \tilde{E}^{a}_j(0, \vec{0}) \tilde{E}^{a}_j(s, \vec{0})\rangle,\label{eq:auv_operator}\\
	\tilde{E}^{a}_i(s, \vec{0}) &= \Omega(s)^\dagger E^{a}_i(s, \vec{0}) \Omega(s),\\
	\Omega(s) &= \text{exp}\left[  -ig \int_{-\infty}^{s} \text{d}s' A_{0}(s', \vec{0}) \right].
\end{align}
$h_{u,v}$ is the singlet or octet Hamiltonian: 
$h_{s,o}=\vec{p}^{2}/M+V_{s,o}$ where $V_{s}= -C_{f}\alpha_{\rm s}(1/a_{0})/r$ and $V_{o}=\alpha_{\rm s}(1/a_{0})/2N_{c}r$, 
with $C_{f}=(N_{c}^{2}-1)/2N_{c} = 4/3$ the Casimir of the fundamental representation, $N_{c}=3$ the number of colors, and $\alpha_{\rm s}(1/a_{0})$ the strong coupling evaluated at the inverse of the Bohr radius $a_{0}$. 

These equations can be written as the master equation
\begin{equation}\label{eq:master_equation}
	\frac{d\rho(t)}{dt} = -i \left[ H, \rho(t) \right] + \sum_{nm} h_{nm} \left( L_{i}^{n} \rho(t) L^{m\dagger}_{i} - \frac{1}{2} \left\{ L^{m\dagger}_{i} L_{i}^{n}, \rho(t) \right\} \right),
\end{equation}
where
\begin{equation}\label{eq:master_equation_rho_and_h}
	\rho(t) = \begin{pmatrix} \rho_{s}(t) & 0 \\ 0 & \rho_{o}(t) \end{pmatrix} ,\text{\quad} H = \begin{pmatrix} h_{s} + \text{Im}(\Sigma_{s}) & 0 \\ 0 & h_{o} + \text{Im}(\Sigma_{o}) \end{pmatrix} ,
\end{equation}
\begin{equation}\label{eq:master_equation_l0_l1}
	L_{i}^{0} = \begin{pmatrix} 0 & 0 \\ 0 & 1 \end{pmatrix}r_{i} \text{, \quad} L_{i}^{1} = \begin{pmatrix} 0 & 0 \\ 0 & \frac{N_{c}^{2}-4}{2(N_{c}^{2}-1)} A_{i}^{oo \dagger} \end{pmatrix},
\end{equation}
\begin{equation}\label{eq:master_equation_l2_l3}
	L_{i}^{2} = \begin{pmatrix} 0 & 1 \\ 1 & 0 \end{pmatrix}r_{i} \text{, \quad} L_{i}^{3} = \begin{pmatrix} 0 & \frac{1}{N_{c}^{2}-1} A_{i}^{os \dagger} \\ A_{i}^{so \dagger} & 0 \end{pmatrix},
\end{equation}
and
\begin{equation}\label{eq:metric_tensor}
	h = \begin{pmatrix} 0 & 1 & 0 & 0 \\ 1 & 0 & 0 & 0 \\ 0 & 0 & 0 & 1 \\ 0 & 0 & 1 & 0 \end{pmatrix}.
\end{equation}
Although eq.~(\ref{eq:master_equation}) is Markovian and trace-preserving, as the matrix $h$ is not completely positive definite, it does not preserve the complete positivity of the density matrix, therefore, it cannot be transformed into Lindblad form.

\section{LO Lindblad equation in \texorpdfstring{$E/T$}{E/T} expansion}
\label{sec4}
In specific temperature regimes, the evolution equation~(\ref{eq:master_equation}) can be rewritten in Lindblad form. In the limit $T\gg E$, the exponentials $A_{i}^{uv}$ which scale like $E/T$, can be written in terms of the transport coefficients $\kappa$ and $\gamma$ \cite{Brambilla:2022ynh}
\begin{equation}
    A_{i}^{uv} = \frac{r_{i}}{2} \left( \kappa - i \gamma \right) + \cdots,
\end{equation}
where $\kappa$ and $\gamma$ are the heavy quarkonium momentum diffusion coefficient and its dispersive counterpart, respectively.
\begin{align}
	\kappa &= \frac{g^{2}}{6N_{c}} \int_{0}^{\infty}\text{d}s\, \left\langle \left\{\tilde{E}^{a}_i(s,\vec{0}), \tilde{E}^{a}_i(0,\vec{0})\right\}\right\rangle ,\label{eq:kappa}\\ 
	\gamma &= -i\frac{g^{2}}{6N_{c}} \int_{0}^{\infty}\text{d}s\, \left\langle \left[\tilde{E}^{a}_i(s,\vec{0}), \tilde{E}^{a}_i (0,\vec{0})\right]\right\rangle.\label{eq:gamma}
\end{align}
At leading order in \( E/T \), the functions \( L^n \) are not linearly independent. Consequently, there exists a rotation of the vector \( (L^0, L^1, L^2, L^3) \) that renders it orthogonal to the eigenspaces corresponding to the negative eigenvalues of the matrix \( h_{nm} \). This enables the master equation to be expressed in Lindblad form.
\begin{equation}\label{eq:lindblad_lo}
\frac{d\rho(t)}{dt} = -i \left[ H, \rho(t) \right] + \sum_{n}\left( C_{i}^{n} \rho(t) C^{n\dagger}_{i} - \frac{1}{2} \left\{ C^{n \dagger}_{i} C_{i}^{n}, \rho(t) \right\} \right),
\end{equation}
with Hamiltonian
\begin{equation}
	H = \begin{pmatrix} h_{s} + \text{Im}(\Sigma_{s}) & 0 \\ 0 & h_{o} + \text{Im}(\Sigma_{o}) \end{pmatrix},
\end{equation}
where
\begin{equation}\label{eq:mass_shift_lo}
	\text{Im}\left( \Sigma_{s} \right) = \frac{r^{2}}{2} \gamma,\quad
	\text{Im}\left( \Sigma_{o} \right) = \frac{N_{c}^{2}-2}{2(N_{c}^{2}-1)} \frac{r^{2}}{2} \gamma,
\end{equation}
and collapse operators
\begin{align}
	C_{i}^{0} =& \sqrt{\frac{\kappa}{N_{c}^{2}-1}} r_{i} \begin{pmatrix} 0 & 1 \\ \sqrt{N_{c}^{2}-1} & 0 \end{pmatrix},\label{eq:c0_lo}\\
	C_{i}^{1} =& \sqrt{\frac{\kappa(N_{c}^{2}-4)}{2(N_{c}^{2}-1)}} r_{i} \begin{pmatrix} 0 & 0 \\ 0 & 1 \end{pmatrix}  \label{eq:c1_lo}.
\end{align}
\section{ NLO Lindblad equation in \texorpdfstring{$E/T$}{E/T} expansion}
\label{sec5}
We use the relation
\begin{equation}\label{eq:nlo_kappa}
	i\frac{g^{2}}{6 N_{c}}\int_{0}^{\infty}\text{d}t \, t \, \Big \langle \tilde{E}^{a}_i(t,\vec{0})\tilde{E}^{a}_i(0,\vec{0}) \Big \rangle = \frac{\kappa}{4T},
\end{equation}
to obtain corrections to eqs.~(\ref{eq:mass_shift_lo})-(\ref{eq:c1_lo}) of order $E/T$ by expanding the exponentials of eq.~(\ref{eq:auv_operator}) and retaining terms up to order $h_{u,v}$.
\begin{align}\label{eq:nlo_medium_interaction}
	A_{i}^{uv} =& \frac{r_{i}}{2} (\kappa - i \gamma) + \kappa \left( -\frac{i p_{i}}{2MT} + \frac{\Delta V_{uv}}{4T} r_{i} \right) + \cdots,
\end{align}
where $\Delta V_{uv}=V_{u}-V_{v}$ is the difference of the $u$ and $v$ (singlet or octet) potentials. With these contributions to $A_{i}^{uv}$, the operators $L_i^1$ and $L_i^3$ in the master equation take the form
\begin{align}
	&L_{i}^{1} = \frac{N_{c}^{2}-4}{2(N_{c}^{2}-1)} \begin{pmatrix} 0 & 0 \\ 0 &  1 \end{pmatrix} \left[ \frac{r_{i}}{2} (\kappa + i \gamma) + \kappa \frac{i p_{i}}{2MT} \right] ,
	\label{eq:Li1NLO}\\
	&\begin{aligned} L_{i}^{3}=
	&\begin{pmatrix} 0 & \frac{1}{N_{c}^{2}-1} \\ 0 & 0 \end{pmatrix} \left[\frac{r_{i}}{2} (\kappa + i \gamma) + \kappa \left( \frac{i p_{i}}{2MT} + \frac{\Delta V_{os}}{4T} r_{i} \right) \right] \\
	&+ \begin{pmatrix} 0 & 0 \\ 1 & 0 \end{pmatrix} \left[\frac{r_{i}}{2} (\kappa + i \gamma) + \kappa \left( \frac{i p_{i}}{2MT} + \frac{\Delta V_{so}}{4T} r_{i} \right) \right],
	\end{aligned}
\end{align}
while $L_i^0$ and $L_i^2$ keep the form given in eqs.~\eqref{eq:master_equation_l0_l1} and~\eqref{eq:master_equation_l2_l3}, respectively.

The master equation above cannot be written in Lindblad form because the functions \( L^n_i \) are linearly independent. However, we can rotate the vector \( (L^0, L^1, L^2, L^3) \) so that only terms of order \( E/T \) have components in the eigenspaces of the negative eigenvalues of \( h_{nm} \). Ignoring these small components is equivalent to neglecting terms of order \( (E/T)^2 \) in the evolution equation. As a result, we can still write a Lindblad equation that remains accurate up to order \( E/T \).
\begin{equation}
\frac{d\rho(t)}{dt} = -i \left[ H, \rho(t) \right] + \sum_{n=0}^1
\left( C_{i}^{n} \rho(t) C^{n\dagger}_{i} - \frac{1}{2} \left\{ C^{n \dagger}_{i} C_{i}^{n}, \rho(t) \right\} \right),
\label{eq:Lindblad}
\end{equation}
with Hamiltonian
\begin{equation}
	H = \begin{pmatrix} h_{s} + \text{Im}(\Sigma_{s}) & 0 \\ 0 & h_{o} + \text{Im}(\Sigma_{o}) \end{pmatrix},
\end{equation}
where
\begin{equation}\label{eq:self_energies_im}
	\text{Im}\left( \Sigma_{s} \right) = \frac{r^{2}}{2} \gamma +\frac{\kappa}{4MT} \{r_{i}, p_{i}\} \text{, \quad} 
	\text{Im}\left( \Sigma_{o} \right) = \frac{N_{c}^{2}-2}{2(N_{c}^{2}-1)} \left( \frac{r^{2}}{2} \gamma +\frac{\kappa}{4MT} \{r_{i}, p_{i}\} \right),
\end{equation}
and collapse operators
\begin{align}
	&\begin{aligned}\label{eq:c0}
	    C_{i}^{0} =& \sqrt{\frac{\kappa}{N_{c}^{2}-1}} \begin{pmatrix} 0 & 1 \\ 0 & 0 \end{pmatrix} \left(r_{i} + \frac{i p_{i}}{2MT} +\frac{\Delta V_{os}}{4T}r_{i} \right) \\ 
	&+ \sqrt{\kappa} \begin{pmatrix} 0 & 0 \\ 1 & 0 \end{pmatrix} \left(r_{i} + \frac{i p_{i}}{2MT} +\frac{\Delta V_{so}}{4T}r_{i} \right),
	\end{aligned}\\
	&C_{i}^{1} = \sqrt{\frac{\kappa(N_{c}^{2}-4)}{2(N_{c}^{2}-1)}} \begin{pmatrix} 0 & 0 \\ 0 & 1 \end{pmatrix} \left(r_{i} + \frac{i p_{i}}{2MT} \right).\label{eq:c1}
\end{align}
For details, see appendix B of \cite{Brambilla:2022ynh}.

\section{NLO Lindblad equation in the reduced spherical representation}
\label{sec6}
As the system is isotropic, we write the Hamiltonian and collapse operators more compactly by using the reduced wave function $u(t,r) \equiv r R(t,r)$, where the three-dimensional wave function for fixed $l$ and $m$ is given by $\psi(t,\vec{r}) = R(t,r) Y_{lm}(\theta,\phi)$. We refer to quantities acting on the reduced wave function as being in the \textit{reduced spherical representation (RSR)} and denote them with an overbar.

\subsection{Jump operators in RSR}
\label{sec6.1}
The Lindblad equation includes six jump operators, which, when expressed in RSR form, are given by:
\begin{eqnarray}
\overline{C}^{\uparrow}_{s\to o} &=& r  - \frac{N_c \alpha_{\rm s}}{8T} + \frac{1}{2 M T} \left( \frac{\partial}{\partial r}  - \frac{l+1}{r} \right) , \label{eq:cupso}\\
\overline{C}^{\downarrow}_{s\to o} &=& r  - \frac{N_c \alpha_{\rm s}}{8T} + \frac{1}{2 M T} \left( \frac{\partial}{\partial r}  + \frac{l}{r} \right) , \\
\overline{C}^{\uparrow}_{o\to s} &=& r  + \frac{N_c \alpha_{\rm s}}{8T} + \frac{1}{2 M T} \left( \frac{\partial}{\partial r}  - \frac{l+1}{r} \right) , \\
\overline{C}^{\downarrow}_{o\to s} &=& r  + \frac{N_c \alpha_{\rm s}}{8T} + \frac{1}{2 M T} \left( \frac{\partial}{\partial r}  + \frac{l}{r} \right) , \\
\overline{C}^{\uparrow}_{o\to o} &=& r + \frac{1}{2 M T} \left( \frac{\partial}{\partial r}  - \frac{l+1}{r} \right) , \\
\overline{C}^{\downarrow}_{o\to o} &=& r  + \frac{1}{2 M T} \left( \frac{\partial}{\partial r}  + \frac{l}{r} \right) . \label{eq:cdownoo}
\end{eqnarray}
In the QTraj code ~\cite{Omar:2021kra}, we multiply all these operators by a factor of \( T \) to ensure they remain regular when \( T = 0 \).

\subsection{Width operators in RSR}
\label{sec6.2}
The width operators in RSR are given by:
\begin{align}
    \overline{\Gamma}_{o\to s}^{\uparrow} &= \frac{\hat\kappa T^3}{N_c^2-1} \frac{l+1}{2l+1} \left[ \left( r +  \frac{N_c \alpha_{\rm s}}{8T} \right)^2 - \frac{3}{2MT} + \frac{\overline{{\cal D}}^2}{(2 M T)^2} - \frac{1}{2 M T} \left( \frac{N_c \alpha_{\rm s}}{4T} \right) \frac{1}{r} \right] , \label{eq:gupos} \\
    \overline{\Gamma}_{o\to s}^{\downarrow} &= \frac{l}{l+1} \overline{\Gamma}_{o\to s}^{\uparrow} \, , \\
    \overline{\Gamma}_{o\to o}^{\uparrow} &=  \hat\kappa T^3 \frac{N_c^2-4}{2(N_c^2-1)} \frac{l+1}{2l+1} \left[ r^2 - \frac{3}{2MT} + \frac{\overline{{\cal D}}^2}{(2 M T)^2} \right] \, , \\
    \overline{\Gamma}_{o\to o}^{\downarrow} &= \frac{l}{l+1}  \overline{\Gamma}_{o\to o}^{\uparrow} \, , \label{eq:gdownoo}
\end{align}
with
\begin{align}
\overline{{\cal D}}^2 = -\frac{\partial^2}{\partial r^2} + \frac{l (l+1)}{r^2} \, .
\end{align}
In this case, up and down transitions are related by $\overline{\Gamma}^\downarrow_{s \to o} / \overline{\Gamma}^\uparrow_{s \to o} = l/(l+1)$.

\subsection{Effective Hamiltonian in RSR}
\label{sec6.3}
The effective Hamiltonian for singlet and octet evolution is defined by $H^{\rm eff}_{s,o} = h_{s,o}  + \text{Im}(\Sigma_{s,o})  - i\Gamma_{s,o}/2$ with $\Gamma_s = \sum_{i \in \{\uparrow,\downarrow\}} \Gamma_{s\rightarrow o}^i$ and $\Gamma_o =  \sum_{i \in \{\uparrow,\downarrow\}} (\Gamma_{o\rightarrow s}^i + \Gamma_{o\rightarrow o}^i$).  
In RSR, the singlet effective Hamiltonian $\overline{H}^{\rm eff}_s$ is given by
\begin{align}
    \text{Re}[\overline{H}^{\rm eff}_s] &=  
    \frac{\overline{\cal D}^2}{M}
    - \frac{C_{f}\, \alpha_{\rm s}}{r} + \frac{\hat\gamma T^3}{2} r^2 +  \frac{\hat\kappa T^2}{4 M} \{r,p_r\} \, , \label{eq:heff1}\\
    \text{Im}[\overline{H}^{\rm eff}_s] &= - \frac{\hat\kappa T^3}{2} \left[ \left( r - \frac{N_c \alpha_{\rm s}}{8T} \right)^2 - \frac{3}{2MT} + \frac{{\overline{\cal D}}^2}{(2MT)^2} + \frac{1}{2MT} \left( \frac{N_c \alpha_{\rm s}}{4 T} \right)  \frac{1}{r}\right] ,
\end{align}
where $p_r = - i \partial_r$.  Similarly, the octet effective Hamiltonian $\overline{H}^{\rm eff}_o$ is given by
\begin{align}
    \text{Re}[\overline{H}^{\rm eff}_o] &=  \frac{\overline{\mathcal{D}}^{2}}{M} + \frac{1}{2N_{c}} \frac{\alpha_{\rm s}}{r} + \frac{N_c^2 - 2}{2(N_c^2 - 1)}\left[ \frac{ \hat\gamma T^3 }{2} r^2 +  \frac{\hat\kappa T^2}{4 M} \{r,p_r\} \right]  , \\
    \text{Im}[\overline{H}^{\rm eff}_o] &= - \frac{\hat\kappa T^3}{2(N_c^2-1)} \left[ \left( r + \frac{N_c \alpha_{\rm s}}{8T} \right)^2 - \frac{3}{2MT} + \frac{{\overline{\cal D}}^2}{(2MT)^2} - \frac{1}{2MT} \left( \frac{N_c \alpha_{\rm s}}{4 T} \right)  \frac{1}{r}\right]  \nonumber \\
&~ \hspace{0.4cm} {-}\frac{\hat\kappa T^3}{4(N_c^2-1)} \left[ r^2  
- \frac{3}{2MT} + \frac{{\overline{\cal D}}^2}{(2MT)^2} 
 \right] ,  \label{eq:heff4}
\end{align}
where $\hat{\kappa} = \kappa / T^3$ and  $\hat{\gamma} = \gamma / T^3$.

\section{QTraj algorithm}
\label{sec7}
The Lindblad equation given in eq.~(\ref{eq:Lindblad}) with the collapse operators of eqs.~\eqref{eq:c0} and \eqref{eq:c1} describes the evolution of the heavy quarkonium reduced density matrix at next-to-leading-order accuracy in the $E/T$ expansion.  
We numerically solve it without \cite{Brambilla:2022ynh} and with \cite{Brambilla:2023hkw} jumps using the quantum trajectories algorithm ~\cite{Brambilla:2021wkt,Brambilla:2020qwo,Omar:2021kra}. The detail description of the NLO quantum trajectories algorithm can be found in \cite{Brambilla:2022ynh}.

\section{Phenomenological Results}
\label{sec8}
\begin{figure}[ht]
\begin{center}
\includegraphics[width=0.48\linewidth]{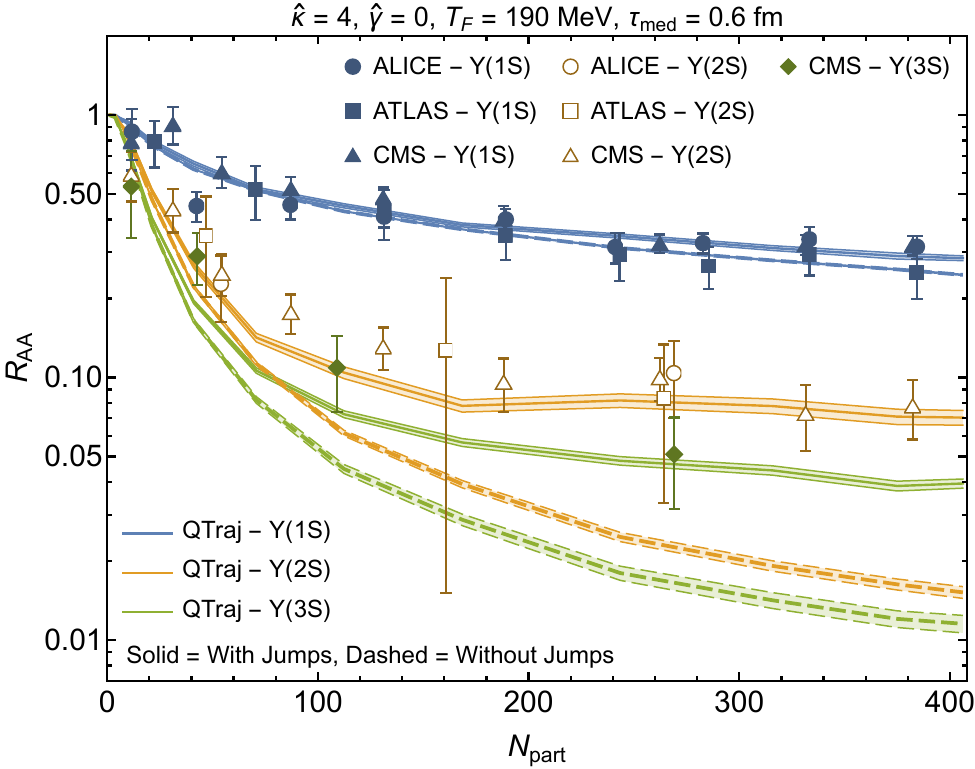} \hspace{3mm}
\includegraphics[width=0.48\linewidth]{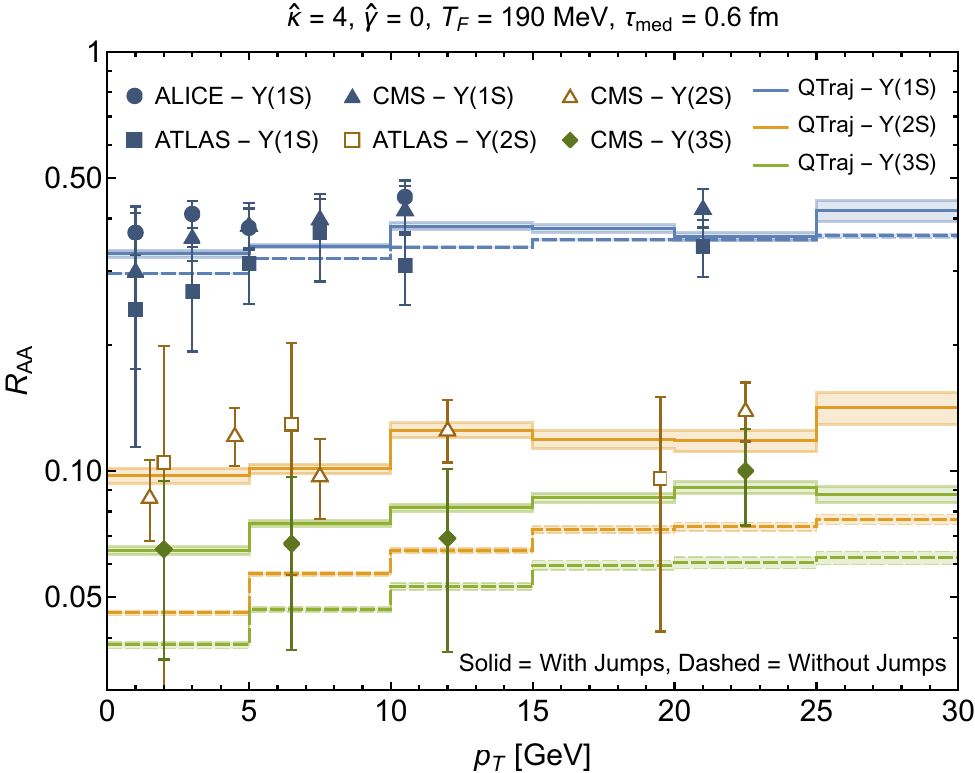}
\end{center}
\caption{The left panel shows nuclear suppression factor $R_{AA}$ of bottomonia states as a function of $N_{\rm part}$. The right panel shows $R_{AA}$ of bottomonia states as a function of $p_T$. The solid and dashed lines indicate results obtained with and without quantum jumps, respectively.  The experimental measurements shown are from the ALICE~\cite{Acharya:2020kls}, ATLAS~\cite{ATLAS5TeV}, and CMS~\cite{Sirunyan:2018nsz,CMS-PAS-HIN-21-007} collaborations. Adapted from \cite{Brambilla:2023hkw}.
}
\label{fig02}
\end{figure}

\begin{figure}[ht]
\begin{center}
\includegraphics[width=0.48\linewidth]{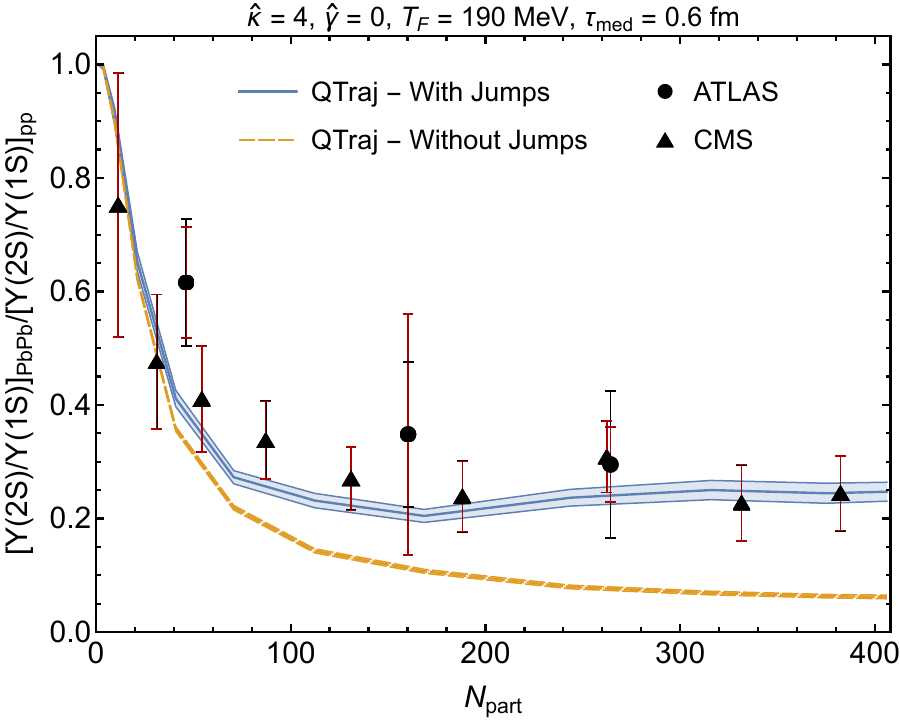} \hspace{3mm}
\includegraphics[width=0.48\linewidth]{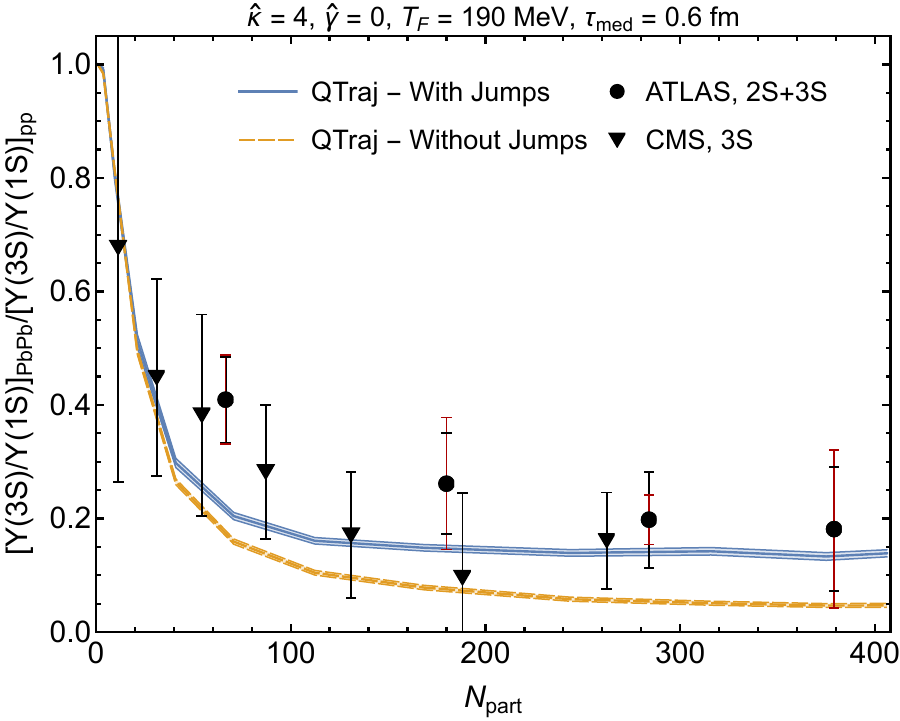}
\end{center}
\caption{The left panel shows the 2S/1S double ratio as a function of $N_{\rm part}$. The right panel shows the 3S/1S double ratio as a function of $N_{\rm part}$. The solid blue lines and dashed orange lines show the QTraj results with and without quantum jumps, respectively. The experimental measurements shown are from the ATLAS~\cite{ATLAS5TeV} and  CMS~\cite{Sirunyan:2018nsz,CMS-PAS-HIN-21-007} collaborations. Adapted from \cite{Brambilla:2023hkw}.
}
\label{fig03}
\end{figure}

\begin{figure}[ht]
\begin{center}
\includegraphics[width=0.48\linewidth]{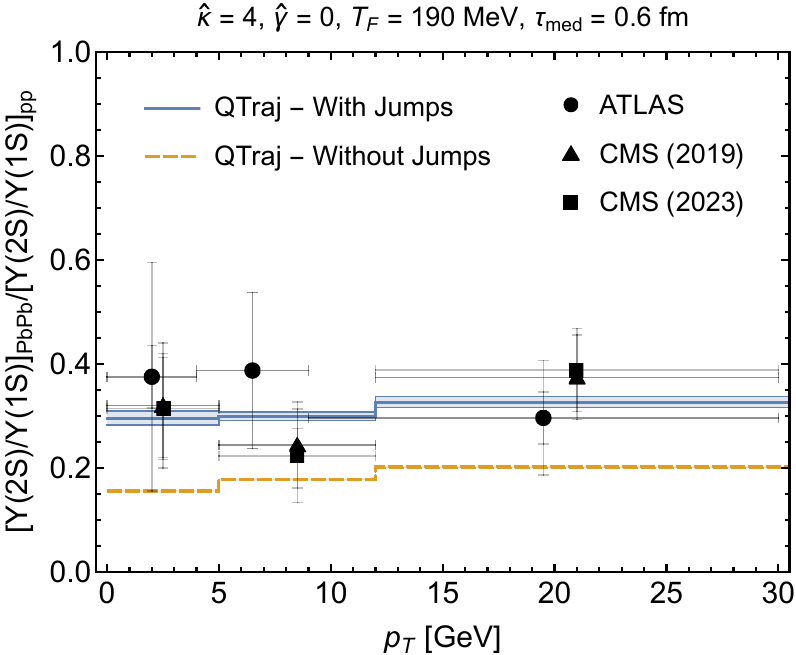} \hspace{3mm}
\includegraphics[width=0.48\linewidth]{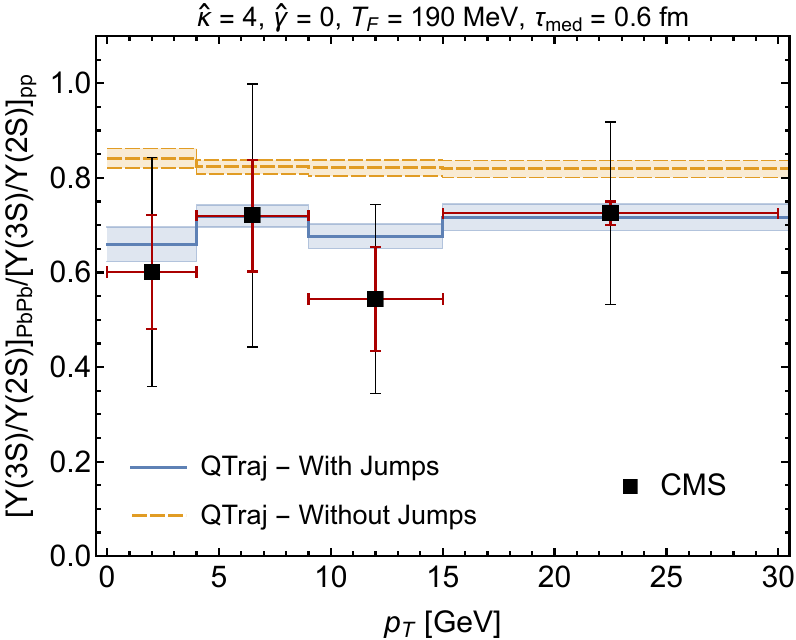}
\end{center}
\caption{The left panel shows the 2S/1S double ratio as a function of $p_T$. The right panel shows the 3S/2S double ratio as a function of $p_T$. Line styles are the same as Fig.~\ref{fig03}. The experimental data were inferred from the results of Refs.~\cite{Sirunyan:2018nsz,CMS-PAS-HIN-21-007}. Adapted from \cite{Brambilla:2023hkw}.
}
\label{fig04}
\end{figure}

We computed the survival probabilities of the $\Upsilon(1S,2S,3S)$ states by taking the ratio of the final and initial overlap probabilities obtained using vacuum Coulomb eigenstates.  
We then took into account the final state feed down by applying a feed down matrix \cite{Brambilla:2020qwo} constructed from data available from the Particle Data Group \cite{Workman:2022ynf}. We showed the nuclear suppression factor $R_{AA}$ of bottomonia states as a function of $N_{\rm part}$ and $p_T$ in the left and right panel of Fig.~\ref{fig02}, respectively. The 2S/1S and the 3S/1S double ratios as a function of $N_{\rm part}$ have been shown in the left and right panels of Fig~\ref{fig03}. These double ratios as a function of $p_T$ have also been shown in Fig~\ref{fig04}. In all these figures, the solid and dashed lines indicate results obtained with and without quantum jumps, respectively.  The experimental measurements shown are from the ALICE~\cite{Acharya:2020kls}, ATLAS~\cite{ATLAS5TeV}, and CMS~\cite{Sirunyan:2018nsz,CMS-PAS-HIN-21-007} collaborations.

\newpage
\section{Summary and Conclusions}
\label{sec9}
We obtain the NLO Lindblad equation in E/T for the evolution of the heavy-quarkonium reduced density matrix using the OQS and pNRQCD approach. We numerically solve the resulting NLO Lindblad equation without and with jumps using the quantum trajectories algorithm. We show that quantum jumps between singlet state and octet states are necessary to understand experimental results for the suppression of both bottomonium ground and excited states. The values of the heavy-quarkonium transport coefficients used are also consistent with recent lattice QCD determinations \cite{Larsen:2019bwy, Bala:2021fkm, Altenkort:2023oms}.
\begin{acknowledgments}
\noindent
N.B. and A.V. acknowledge support by the DFG cluster of excellence ORIGINS funded by the Deutsche Forschungsgemeinschaft under Germany's Excellence Strategy - EXC-2094-390783311.
M.\'{A}.E was supported by European Research Council project ERC-2018-ADG-835105 YoctoLHC, by the Maria de Maetzu excellence program under projects CEX2020-001035-M and CEX2019-000918-M, the Spanish Research State Agency under projects PID2020-119632GB-I00 and PID2019-105614GB-C21, the Xunta de Galicia (Centro singular de investigaci\'on de Galicia accreditation 2019-2022; European Union ERDF). A.I. and M.S. were supported by U.S. Department of Energy Award No.~DE-SC0013470. M.S. also thanks the Ohio Supercomputer Center under the auspices of Project No.~PGS0253.  P.V.G. was supported by the U.S. Department of Energy Award No.~DE-SC0019095.
P.V.G. is grateful for the support and hospitality of the Fermilab theory group.
\end{acknowledgments}

\bibliographystyle{JHEP}
\bibliography{ConfAus.bib}

\providecommand{\href}[2]{#2}\begingroup\raggedright\begin{thebibliography}{10}

\bibitem{Brambilla:2022ynh}
N.~Brambilla, M.~A. Escobedo, A.~Islam, M.~Strickland, A.~Tiwari, A.~Vairo et~al., \emph{{Heavy quarkonium dynamics at next-to-leading order in the binding energy over temperature}}, \href{https://doi.org/10.1007/JHEP08(2022)303}{\emph{JHEP} {\bfseries 08} (2022) 303} [\href{https://arxiv.org/abs/2205.10289}{{\ttfamily 2205.10289}}].

\bibitem{Brambilla:2023hkw}
N.~Brambilla, M.~A. Escobedo, A.~Islam, M.~Strickland, A.~Tiwari, A.~Vairo et~al., \emph{{Regeneration of bottomonia in an open quantum systems approach}}, \href{https://doi.org/10.1103/PhysRevD.108.L011502}{\emph{Phys. Rev. D} {\bfseries 108} (2023) L011502} [\href{https://arxiv.org/abs/2302.11826}{{\ttfamily 2302.11826}}].

\bibitem{Matsui:1986dk}
T.~Matsui and H.~Satz, \emph{{$J/\psi$ suppression by quark-gluon plasma formation}}, \href{https://doi.org/10.1016/0370-2693(86)91404-8}{\emph{Phys. Lett.} {\bfseries B178} (1986) 416}.

\bibitem{Laine:2006ns}
M.~Laine, O.~Philipsen, P.~Romatschke and M.~Tassler, \emph{{Real-time static potential in hot QCD}}, \href{https://doi.org/10.1088/1126-6708/2007/03/054}{\emph{JHEP} {\bfseries 03} (2007) 054} [\href{https://arxiv.org/abs/hep-ph/0611300}{{\ttfamily hep-ph/0611300}}].

\bibitem{Brambilla:2013dpa}
N.~Brambilla, M.~A. Escobedo, J.~Ghiglieri and A.~Vairo, \emph{{Thermal width and quarkonium dissociation by inelastic parton scattering}}, \href{https://doi.org/10.1007/JHEP05(2013)130}{\emph{JHEP} {\bfseries 05} (2013) 130} [\href{https://arxiv.org/abs/1303.6097}{{\ttfamily 1303.6097}}].

\bibitem{Brambilla:2008cx}
N.~Brambilla, J.~Ghiglieri, A.~Vairo and P.~Petreczky, \emph{Static quark-antiquark pairs at finite temperature}, \href{https://doi.org/10.1103/PhysRevD.78.014017}{\emph{Phys. Rev.} {\bfseries D78} (2008) 014017} [\href{https://arxiv.org/abs/0804.0993}{{\ttfamily 0804.0993}}].

\bibitem{Akamatsu:2011se}
Y.~Akamatsu and A.~Rothkopf, \emph{{Stochastic potential and quantum decoherence of heavy quarkonium in the quark-gluon plasma}}, \href{https://doi.org/10.1103/PhysRevD.85.105011}{\emph{Phys. Rev.} {\bfseries D85} (2012) 105011} [\href{https://arxiv.org/abs/1110.1203}{{\ttfamily 1110.1203}}].

\bibitem{Brambilla:2016wgg}
N.~Brambilla, M.~A. Escobedo, J.~Soto and A.~Vairo, \emph{{Quarkonium suppression in heavy-ion collisions: an open quantum system approach}}, \href{https://doi.org/10.1103/PhysRevD.96.034021}{\emph{Phys. Rev.} {\bfseries D96} (2017) 034021} [\href{https://arxiv.org/abs/1612.07248}{{\ttfamily 1612.07248}}].

\bibitem{Sharma:2019xum}
R.~Sharma and A.~Tiwari, \emph{{Quantum evolution of quarkonia with correlated and uncorrelated noise}}, \href{https://doi.org/10.1103/PhysRevD.101.074004}{\emph{Phys. Rev. D} {\bfseries 101} (2020) 074004} [\href{https://arxiv.org/abs/1912.07036}{{\ttfamily 1912.07036}}].

\bibitem{Blaizot:2015hya}
J.-P. Blaizot, D.~De~Boni, P.~Faccioli and G.~Garberoglio, \emph{{Heavy quark bound states in a quark\textendash{}gluon plasma: Dissociation and recombination}}, \href{https://doi.org/10.1016/j.nuclphysa.2015.10.011}{\emph{Nucl. Phys. A} {\bfseries 946} (2016) 49} [\href{https://arxiv.org/abs/1503.03857}{{\ttfamily 1503.03857}}].

\bibitem{Yao:2018nmy}
X.~Yao and T.~Mehen, \emph{{Quarkonium in-medium transport equation derived from first principles}}, \href{https://doi.org/10.1103/PhysRevD.99.096028}{\emph{Phys. Rev. D} {\bfseries 99} (2019) 096028} [\href{https://arxiv.org/abs/1811.07027}{{\ttfamily 1811.07027}}].

\bibitem{Katz:2015qja}
R.~Katz and P.~B. Gossiaux, \emph{{The Schrödinger–Langevin equation with and without thermal fluctuations}}, \href{https://doi.org/10.1016/j.aop.2016.02.005}{\emph{Annals Phys.} {\bfseries 368} (2016) 267} [\href{https://arxiv.org/abs/1504.08087}{{\ttfamily 1504.08087}}].

\bibitem{Gross:2022hyw}
F.~Gross et~al., \emph{{50 Years of Quantum Chromodynamics}},  \href{https://arxiv.org/abs/2212.11107}{{\ttfamily 2212.11107}}.

\bibitem{Brambilla:2017zei}
N.~Brambilla, M.~A. Escobedo, J.~Soto and A.~Vairo, \emph{{Heavy quarkonium suppression in a fireball}}, \href{https://doi.org/10.1103/PhysRevD.97.074009}{\emph{Phys. Rev.} {\bfseries D97} (2018) 074009} [\href{https://arxiv.org/abs/1711.04515}{{\ttfamily 1711.04515}}].

\bibitem{Omar:2021kra}
H.~B. Omar, M.~A. Escobedo, A.~Islam, M.~Strickland, S.~Thapa, P.~Vander~Griend et~al., \emph{{QTRAJ 1.0: A Lindblad equation solver for heavy-quarkonium dynamics}}, \href{https://doi.org/10.1016/j.cpc.2021.108266}{\emph{Comput. Phys. Commun.} {\bfseries 273} (2022) 108266} [\href{https://arxiv.org/abs/2107.06147}{{\ttfamily 2107.06147}}].

\bibitem{Brambilla:2021wkt}
N.~Brambilla, M.~A. Escobedo, M.~Strickland, A.~Vairo, P.~Vander~Griend and J.~H. Weber, \emph{{Bottomonium production in heavy-ion collisions using quantum trajectories: Differential observables and momentum anisotropy}}, \href{https://doi.org/10.1103/PhysRevD.104.094049}{\emph{Phys. Rev. D} {\bfseries 104} (2021) 094049} [\href{https://arxiv.org/abs/2107.06222}{{\ttfamily 2107.06222}}].

\bibitem{Brambilla:2020qwo}
N.~Brambilla, M.~A. Escobedo, M.~Strickland, A.~Vairo, P.~Vander~Griend and J.~H. Weber, \emph{{Bottomonium suppression in an open quantum system using the quantum trajectories method}}, \href{https://doi.org/10.1007/JHEP05(2021)136}{\emph{JHEP} {\bfseries 05} (2021) 136} [\href{https://arxiv.org/abs/2012.01240}{{\ttfamily 2012.01240}}].

\bibitem{Workman:2022ynf}
{\scshape Particle Data Group} collaboration, \emph{{Review of Particle Physics}}, \href{https://doi.org/10.1093/ptep/ptac097}{\emph{PTEP} {\bfseries 2022} (2022) 083C01}.

\bibitem{Acharya:2020kls}
{\scshape ALICE} collaboration, \emph{{$\Upsilon$ production and nuclear modification at forward rapidity in Pb\textendash{}Pb collisions at $s_{NN}=5.02$ TeV}}, \href{https://doi.org/10.1016/j.physletb.2021.136579}{\emph{Phys. Lett. B} {\bfseries 822} (2021) 136579} [\href{https://arxiv.org/abs/2011.05758}{{\ttfamily 2011.05758}}].

\bibitem{ATLAS5TeV}
{ATLAS Collaboration}, \emph{{Production of $\varUpsilon(\textrm{nS})$ mesons in Pb+Pb and $pp$ collisions at 5.02 TeV}},  5, 2022.

\bibitem{Sirunyan:2018nsz}
{\scshape CMS} collaboration, \emph{{Measurement of nuclear modification factors of $\Upsilon$(1S), $\Upsilon$(2S), and $\Upsilon$(3S) mesons in PbPb collisions at $\sqrt{s_{_\mathrm{NN}}} =$ 5.02 TeV}}, \href{https://doi.org/10.1016/j.physletb.2019.01.006}{\emph{Phys. Lett. B} {\bfseries 790} (2019) 270} [\href{https://arxiv.org/abs/1805.09215}{{\ttfamily 1805.09215}}].

\bibitem{CMS-PAS-HIN-21-007}
{CMS Collaboration}, \emph{{Observation of the $\Upsilon\textrm{(3S)}$ meson and sequential suppression of $\Upsilon$ states in PbPb collisions at $\sqrt{\mathrm{s_{NN}}}=5.02~\mathrm{TeV}$}},  tech. rep., CERN, Geneva, 2022.

\bibitem{Larsen:2019bwy}
R.~Larsen, S.~Meinel, S.~Mukherjee and P.~Petreczky, \emph{{Thermal broadening of bottomonia: Lattice nonrelativistic QCD with extended operators}}, \href{https://doi.org/10.1103/PhysRevD.100.074506}{\emph{Phys. Rev. D} {\bfseries 100} (2019) 074506} [\href{https://arxiv.org/abs/1908.08437}{{\ttfamily 1908.08437}}].

\bibitem{Bala:2021fkm}
{\scshape HotQCD} collaboration, \emph{{Static quark-antiquark interactions at nonzero temperature from lattice QCD}}, \href{https://doi.org/10.1103/PhysRevD.105.054513}{\emph{Phys. Rev. D} {\bfseries 105} (2022) 054513} [\href{https://arxiv.org/abs/2110.11659}{{\ttfamily 2110.11659}}].

\bibitem{Altenkort:2023oms}
L.~Altenkort, O.~Kaczmarek, R.~Larsen, S.~Mukherjee, P.~Petreczky, H.-T. Shu et~al., \emph{{Heavy Quark Diffusion from 2+1 Flavor Lattice QCD}},  \href{https://arxiv.org/abs/2302.08501}{{\ttfamily 2302.08501}}.

\end{thebibliography}\endgroup

\end{document}